\documentclass{article}
\usepackage{spconf}%,amsmath,graphicx}
\usepackage{color}
\usepackage[pdftex]{graphicx}
\DeclareGraphicsExtensions{.pdf,.jpeg,.png}
\graphicspath{{./figures/}}

\usepackage[cmex10]{amsmath}
\usepackage{amssymb, latexsym}
\usepackage{amsfonts}
\usepackage{amsthm}
\usepackage{mathrsfs}    % special font for L-operator $\mathscr{L}$
\usepackage{soul}

\newcommand{\be}{\begin{equation}}
\newcommand{\ee}{\end{equation}}
\def\bal#1\eal{\begin{align}#1\end{align}}

\newcounter{saveeqn}

\title{Speech to Speech Synthesis for Voice Impersonation}

\name{Bjorn Johnson and Jared Levy \thanks{Both authors contributed equally to this work}}
\address{University of California San Diego, La Jolla, CA 92093-0238 \\
Department of Computer Science and Engineering \\
\{bljohnso, jalevy\}@ucsd.edu}

\newcommand{\squeeze}{\vspace*{-0ex}}

\date{January 2020} % This won't show in spconf by default, so use the line below

\usepackage{fancyhdr}
\fancypagestyle{firstpage}{
    \fancyhf{} 
    
    \fancyhead[L]{\scriptsize Original manuscript dated: April 2020}
    \fancyhead[R]{\scriptsize \thepage} % Page number in top right
}
\begin{document}
\maketitle
% \thispagestyle{myheadings}
% \markboth{Original manuscript dated: April 2020}{Original manuscript dated: April 2020}
\thispagestyle{firstpage}

\section{Abstract}
\squeeze
Numerous models have shown great success in the fields of speech recognition as well as speech synthesis, but models for speech to speech processing have not been heavily explored. We propose Speech to Speech Synthesis Network (STSSN), a model based on current state of the art systems that fuses the two disciplines in order to perform effective speech to speech style transfer for the purpose of voice impersonation. We show that our proposed model is quite powerful, and succeeds in generating realistic audio samples despite a number of drawbacks in its capacity. We benchmark our proposed model by comparing it with a generative adversarial model which accomplishes a similar task, and show that ours produces more convincing results.

{\keywords speech recognition, speech synthesis, GAN, mean opinion score}

\squeeze
\section{Introduction}
\squeeze
Generating realistic voice impersonations is a field which has not received much focus in research, but we believe it has a wide variety of potential benefits and applications. A solution to this problem could be used for identity protection in interactive online settings, overcoming language barriers and speech impediments to ease communication, or increasing diversity of character voices in video games and media without the need to hire a large number of voice actors. At the same time, however, the dangers of such a technology must be considered. Convincing telemarketer scams, falsified audio recordings, and other such malicious applications exist, but we believe that enough positive use cases exist to justify the creation of such a system.

The proposed model takes as input two audio clips, one of a source speaker and a second of a target speaker, and generates a new audio clip with the same language content as the first, but with the voice and mannerisms of the second.

We benchmarked our model by additionally implementing CycleGAN, a known style transfer generative adverserial network, applied to audio spectrograms. The input is an audio clip of a source speaker (which is preprocessed) whose phrase we want to capture as well as an audio clip of the target speaker whose style we want to capture, then, the output is an audio clip of the target speaker saying the source's phrase. The statistic most commonly used to analyze results of this problem is the mean opinion score --- an average score composed from the ratings of real humans; the ratings reflect how realistic someone believes the produced speech is.

CycleGAN mostly follows the normal generative adverserial network technique in which we define a competition between a generator network and a discriminator network, however, for CycleGAN, there are two generators and discriminators. The competition relies on the generator being able to generate results (audio spectrograms of speech in our case) as similar to the training set as possible, while the discriminator learns to be able to discriminate between the generated data and the training data. In CycleGAN, one generator generates A to B, and another is B to A, while each generator also is associated with its own discriminator. We see however, that the GANs model is not very robust to noise and the results are ok but not optimal.
We attempt to get around this issue by combining multiple large pretrained networks that are used for speech recognition with large pretrained networks that synthesize speech from text. Using these state of the art deep neural networks, we can notice better results.

\squeeze
\section{Related Work}
\squeeze
The current approaches to voice impersonation are mostly parametric systems, which heuristically tune aspects of the audio such as frequency and speed, and concatenative systems, which slice up audio samples and stitch phonemes together to generate full speech. These systems are severely lacking in their capacity to modify audio files, and generate highly artificial samples.

In terms of machine learning approaches, there have been a number of different attempts to solve this problem, but none have shown generalizable and realistic results. As far as we can tell, a model that solves this exact problem has only been implemented by a number of startups which haven’t shared details about their models or training data. There are, however, a massive number of systems which perform incredibly similar tasks, such that we were able to make a few modifications to existing networks and concatenate them to create a functional speech-to-speech synthesis network. Both Google \cite{jia2018transfer} and Baidu \cite{arik2018neural} have developed text-to-speech synthesis networks which use “speaker profile” encodings to generate audio which appears to have actually been spoken by their target speaker. Other approaches exist as well, a Carnegie Mellon group \cite{gao2018voice} approached the problem using GANs (generative adversarial networks), training one network to detect fake voices and another to produce a voice from text, forcing the generative network to synthesize natural sounding speech. A Stanford group \cite{verma2018neural} tried yet another approach, implementing neural style transfer on spectrograms in a similar fashion to popular art generation techniques, and found good results. Some even tried more direct conversion techniques, \cite{6891242} used a deep neural net to convert from one speaker’s “spectral envelope” to another’s, transforming instantaneous features such as pitch.

We take inspiration for STSSN from the Tacotron2 model \cite{shen2017natural}, which shows that a recurrent neural network architecture with location sensitive attention can produce incredibly realistic speech samples given only a text sequence as an input. This model still has state of the art performance years after it's creation, but can only generate speech samples in a single voice. We also take inspiration from Baidu's DeepSpeech model \cite{hannun2014deep}, which shows great success in using a recurrent architecture to produce a text transcription given an audio file as input. Our fully formed model is the concatenation of both of these models, with the insertion of a style embedding for the target speaker to the decoding stage, expanding the power of our model to be able to generate speech in multiple different voices without needing to be retrained.

\squeeze
\section{Dataset Details}
\squeeze
For the STSSN we use a subset of the LibriSpeech dataset \cite{7178964}, which is a corpus of 1000 hours of audiobook readings with transcriptions. The dataset features hundreds of different speakers, both male and female with a variety of accents, making the dataset perfect for research into style transfer. Each audio clip is relatively short, up to a few tens of seconds. We used this dataset to train our speech to text model, and utilized a pretrained implementation of Tacotron2 for the decoder. We trained our network on the train-clean-100 subset of the dataset, composed of approximately 100 hours of speech, with about 25 minutes of content per speaker. We validated our model using the test-clean subset, composed of roughly 5.4 hours of speech with 8 minutes per speaker. This data is preprocessed by conversion to a mel spectrogram, using a 128 dimensional mel filterbank and a sample rate of 16kHz. These spectrograms are the actual inputs and outputs of our model.

For the CycleGAN benchmark, we implemented the plethora of male and female audio clips provided by the voice conversion challenge in 2016 titled VCC 2016 Dataset \cite{vcc2016}. This data, split up into 1620 training samples and 108 validation samples, is preprocessesd by first breaking down each spectrogram into its f0 signal, which is the fundamental frequency in speech, its spectral envelope, which is how the amplitude of the spectrogram changes, and its aperiodocity. From there we develop encodings of the spectral envelopes and use the f0 signal to analyze pitch statistics of the audio --- using logarithm gaussian normalization. The encoded spectral envelopes are then passed in as inputs to the network.

This large of a dataset provides many possible predictive tasks beyond ours such as predicting gender of a speaker, speech recognition, and speech synthesis. Our decision to implement speech synthesis with voice style transfer stems from the lack of research already done. Speech recognition and speech synthesis especially has already had a lot of work done and large companies have already produced exciting results.

\squeeze
\section{Methods}
\begin{figure}[h]
    \center{\includegraphics[scale=0.3]{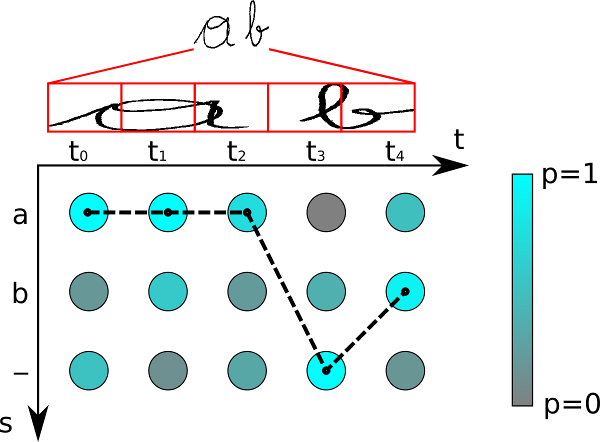}}
    \caption{\label{fig:CTC} Network output for use with CTC loss (dashed line represents best output label), taken from \cite{ctc}}
\end{figure}

\squeeze
As discussed above, we built STSSN off of a collection of pre-existing networks. We utilize a modified DeepSpeech architecture similar to that described in \cite{hannun2014deep} to generate text transcriptions of our content audio files, trained using the CTC loss function. This loss mechanism operates on a network in which the output is a 2D matrix, as shown in Figure \ref{fig:CTC}, with one axis representing timesteps in the sequence and the other representing alphabet labels. The entries correspond to the probability of a specific character being uttered at each timestep. This output is then scored by comparing the output to the true label, at every possible alignment in the sequence. This treats all outputs that differ only by alignment as equivalent, and has been shown to work exceedingly well to decode a high dimensional sequence to a lower dimensional label, in our case a text transcription.

For the target speaker representation, we utilize a speaker encoding network, taken from \cite{corentin2019Code}, which was trained on a speaker discrimination task and shown to create high quality representations of speaker styles. This model uses a stack of LSTM layers followed by a projection to 256 dimensions, generating a fixed dimensional speaker encoding givent an arbitrary length audio sequence input. It is optimized to maximize the cosine similarity of multiple embeddings of the same speaker, while simultaneously maximizing distance between different speakers in the output space.

\begin{figure}[h]
    \center{\includegraphics[scale=0.5]{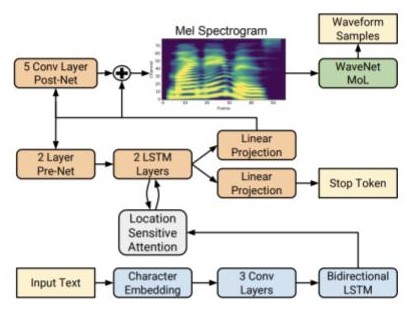}}
    \caption{\label{fig:Tacotron} Tacotron2 Architecture}
\end{figure}

We feed this transcript through an implementation of the Tacotron2 network, shown graphially in Figure \ref{fig:Tacotron}, concatenating the style encoding with the character embedding to synthesize a spectrogram representing a voice with a specific speaker style. This model operates by replacing every character in the sequence with a high dimensional learnable embedding, feeding this embedding to a stack of convolutional layers, and finally a bidirectional LSTM layer to generate an encoding for the input. At this stage, the style encoding is concatenated with the sequence encoding at every timestep, and fed into a location sensitive attention layer to generate a fixed length context vector for each decoder step. The network feeds each decoder output step through a small linear prenet, concatenates that output with the current attention context vector, feeds this into a unidirectional LSTM stack, and uses a linear projection to generate the next output step. This model is trained using text and spectrogram pairs, with a mean squared error loss function on the output spectrogram. Our motivation for this approach comes from the wealth of prior research summarized above, we believed it would be the most practical to implement, as we found open source github repositories for the encoding network as well as the Tacotron implementation, and would have a high potential for end to end success.

Now let's look at the main optimization objective for the CycleGAN.

\includegraphics[scale=0.5]{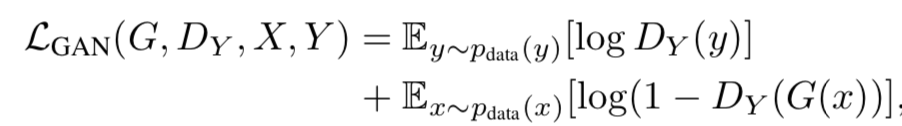}\\
The function above (from \cite{CycleGAN}) is the adverserial loss function which is part of the overall loss in which CycleGAN trains on. The G is the generator function/model that generates results as close to domain Y as possible. D is the discriminator function/model whose goal is to discriminate between data generated by G and actual examples, y. The generator model tries to minimize this loss function while the discriminator tries to maximize it. We notice that if the discriminator is fooled by G, then log(1 - D(G(x)) will decrease and as a result the loss would decrease --- which goes against the discriminators goal. From the other perspeective, if the G fools D and the loss goes down, then G is successfully minimizing the loss and improving. 

\includegraphics[scale=0.5]{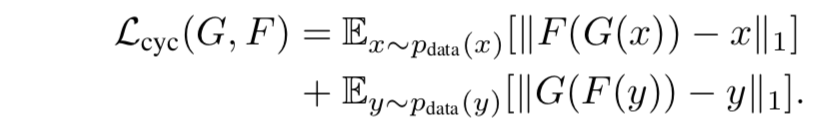}\\
The next part of the CycleGAN's loss is the cycle consistency loss (also from \cite{CycleGAN}). In order to increase robustness, we want our algorithm to be able to generate an audio file and also be able to deconstruct it back into the inputs. This is why we have two generators, G and F. Generator G produces the style transferred result where F's goal is to take in the style transferred result and produce a normal result. Thus, we notice how the formula uses the style transferred result (G(x)) as an input to F, and records its difference with the original non-style transferred data point. The second part of the formula is more familiar in context with our problem. We want to take a normal input and output a new style transferred result that is very similar to how the dataset is.

\includegraphics[scale=0.5]{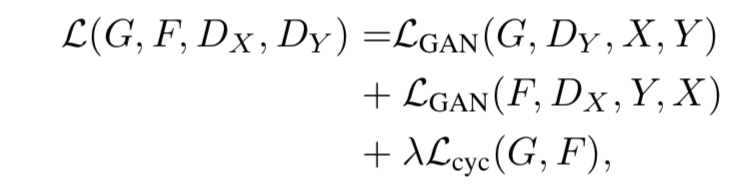}\\
We can define our total loss as the sum of the adverserial losses of our two generators/discriminators with the cycle consistency loss. Lambda is a hyperparameter for the importance of cycle consistency.

\squeeze
\section{Experiments/Results/Discussion}
\squeeze

\begin{figure}[h]
    \center{\includegraphics[scale=0.6]{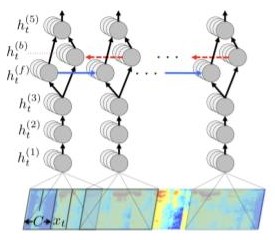}}
    \caption{\label{fig:DeepSpeech} DeepSpeech Architecture}
\end{figure}

We'll describe the DeepSpeech model a bit more in depth, as this was the component most of our effort went into. Summarized graphically in Figure \ref{fig:DeepSpeech}, this model consists of a single convolutional layer with 32 filters for basic feature extraction, followed by 3 residual blocks, each composed of 2 convolutional layers with 32 filters and layer normalization. This feeds into a fully connected layer which consumes the vertical dimension and all of the channels, and outputs 256 features per timestep in the sequence. The resulting sequence is run through 4 bidirectional GRU layers, each with 512 hidden units, one more fully connected layer with 512 hidden units and a GELU activation function, and finally a fully connected layer with 29 hidden units. We trained this model using the CTC Loss Function on a 29 character alphabet in conjunction with a greedy decoder. We used a batch size of 8, as this was the highest our GPU could handle without running out of memory, and trained to convergence after roughly 70 epochs of our dataset. We used a one-cycle learning rate scheduler, which linearly increased learning rate to a max of 5e-4, and then linearly decayed. Training took roughly 48 hours on a single GPU, so it was difficult for us to experiment with many of our network's parameters, but we did try tweaking a number of hyperparameters and comparing results. For example, research suggested using 5 GRU layers each with 1024 hidden units, but we found that cutting this down to 4 layers with only 512 units each still produced decent results. We cut down on the capacity of our model in a number of places to speed up training and reduce model size, as the fidelity of our text transcription needed not be perfect. Since we fed the transcription back into Tacotron2, minor misspellings and word errors often did not greatly impact the quality of the final result, likely because misspellings would typically have similar pronunciations to the correct word. More experimentation in this area would be valuable, and thus will be described more in the future work section.

\begin{figure}[h]
    \center{\includegraphics[scale=0.26]{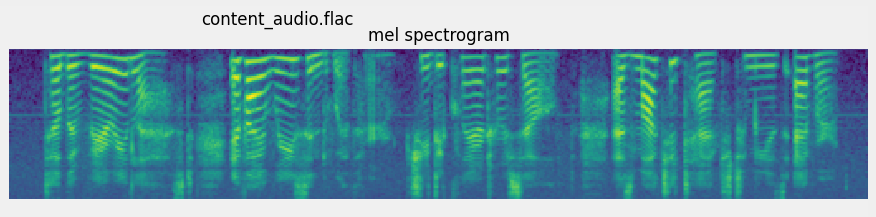}}\\
    \center{\includegraphics[scale=0.26]{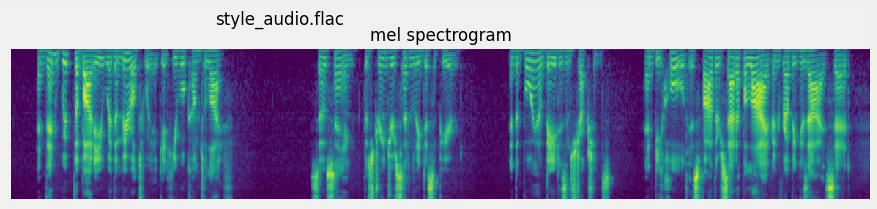}}\\
    \center{\includegraphics[scale=0.26]{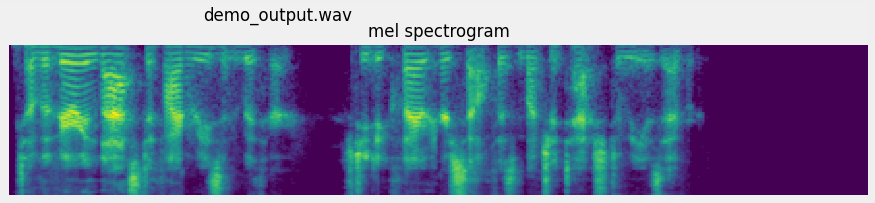}}\\
    \caption{\label{fig:Specs} Spectrograms of content (top), style (mid), and output (bottom) audio files}
\end{figure}

We leave the Tacotron2 implementation unchanged, exactly to the specifications set forth in \cite{shen2017natural}, using the pretrained implementation provided in \cite{corentin2019Code}. This produces a mel spectrogram, which we process into an audio file using a vocoder contained in the same repository. We see some example results of our full network in Figure \ref{fig:Specs}, and can visually compare the spectrograms to get an idea of how our network functions. Listening to the audio files is by far the best way to understand the success of our model, so please refer to the demo section of our presentation for that, but you can also qualitatively see results through the spectrograms of the three audio files. That the output spectrogram contains similar features to the content spectrogram (pay attention to the patterns, the horizontal stripes and vertical blanks are preserved), while having an overall appearance more similar to the style spectrogram (less haze, similar blotchiness).

\begin{figure}[h]
    \center{\includegraphics[scale=0.4]{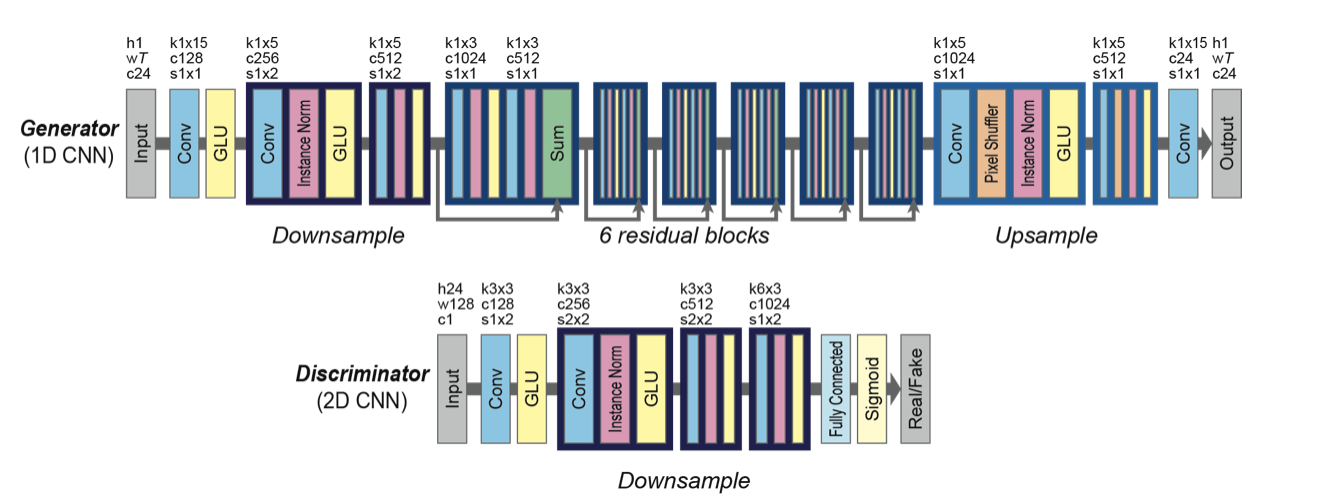}}\\
    \caption{\label{fig:CycleGANARchitecture} CycleGAN Architecture}
\end{figure}

In terms of hyperparameters chosen for the CycleGAN, we can see the architecture of the model in Figure \ref{fig:CycleGANARchitecture}. This architecture stems from the architecture seen in \cite{CycleGAN}.  Since our problem is heavily recurrent, we use the gated linear unit (GLU) activation. This neuron activation performs well in recurrent tasks by learning when to pass information down the sequence and when not to. Furthermore, the GAN's network uses a minibatch size of 1 (1 works the best out of the multiple we tried). Although this is unusual in most deep networks, the small batch size allows our model to perform better, but takes longer in training. This hyperparameter was chosen because the dataset wasn't very large.

Although the general adverserial loss function has the discriminator attempting to maximize, it is actually trivial, as long as the generator and discriminator have opposite goals (e.g. generator maximizes and discriminator minimizes or vice versa). We see this in our code heavily influenced from \cite{CycleGANvoiceCode} in which the losses with respect to iteration is shown below.

\begin{figure}[h]
    \center{\includegraphics[scale=0.35]{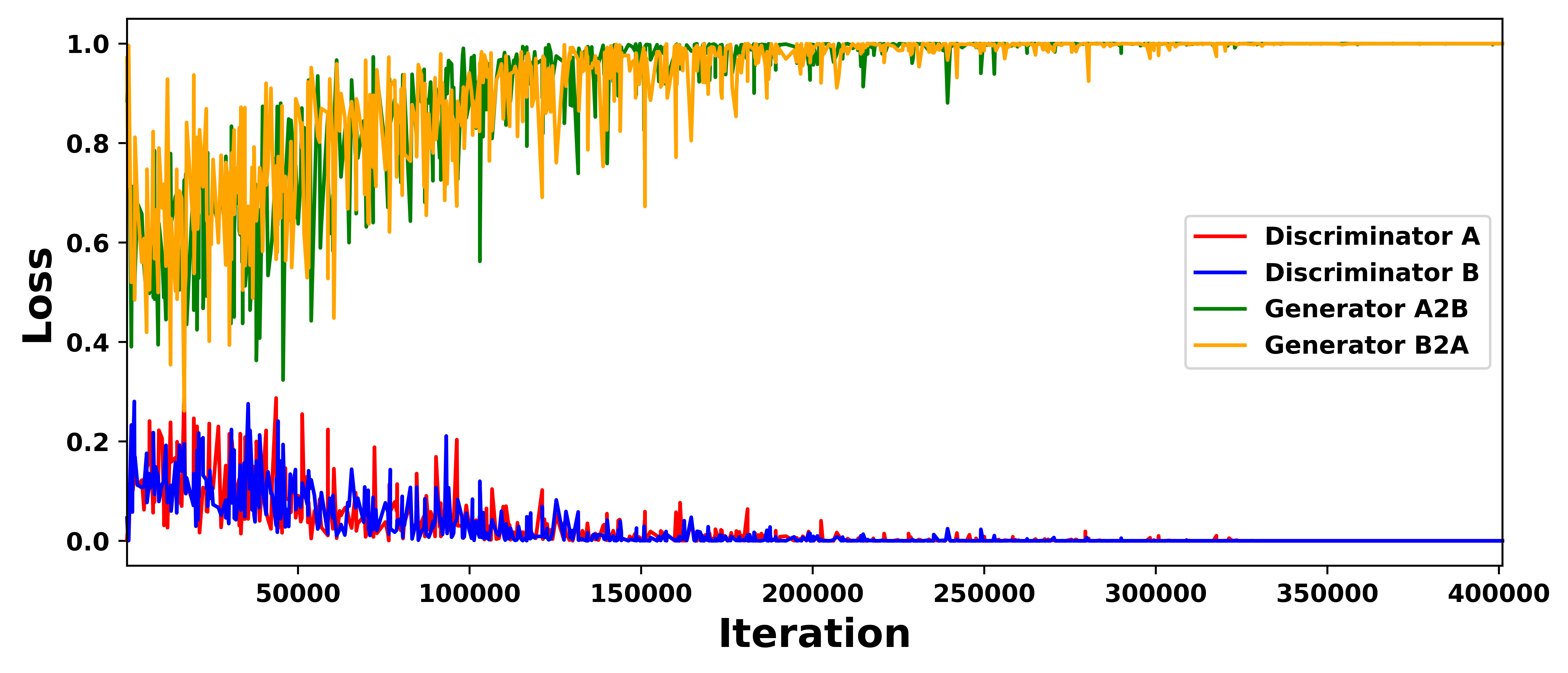}}\\
    \caption{\label{fig:GANLoss} GAN model losses}
\end{figure}

We see that here the generators attempt to maximize loss while our discriminators attempt to minimize the CycleGAN loss explained in the Methods portion. Either way, we can see that the results are very promising.

However, after analysis of the results of both the CycleGANs and the STSSN, we can noticibly tell that although CycleGAN was good, it seemed to overfit, while the STSSN produced very interesting results and thus, we spent more timing tuning the hyperparameters for the STSSN.

\squeeze
\section{Conclusion/Future Work}
\squeeze
We present a concatenation of a high quality speech to text recognition model with a state of the art text to speech synthesis model conditioned to generate speech in a variety of styles, and show that this conglomerate model has high potential in use as a speech impersonation network. We also implement a rivalling technique, CycleGAN, and compare outputs to see that the conglomerate model is capable of producing much more realistic speech.

Our results were significantly better than we had initially expected, and thus we are looking to the future to continue this work and develop an even better voice impression system. Our current approach is limited by the fact that the output of our encoding network must be decoded to raw text, then encoded again to be run through the synthesis network. We are in the process of rebuilding our model as a single, end to end model rather than a concatenation of separate models, by removing the decoder from the first network, and the encoder from the second, with the intention of generating higher quality embeddings for the spectrograms and avoiding the text bottleneck. We predict that this change will preserve significantly more information from the input audio, making it easier for the network to generate highly realistic samples. In addition, this change would reduce the size of the overall model, resulting in reduced training times and more potential for hyperparameter experimentation. This is, however, a massive undertaking and is beyond the scope of this course and research project due to time constraints. In addition to rewriting our code as a single model, we would need to establish a loss function that forces the output to resemble both the content of the first input, and the style of the second. This would likely be done through a triplet loss function, by comparing the distance of the output to the content input as well as the style input using mean squared error of the corresponding embeddings, and summing these losses together.

\bibliographystyle{unsrt}
\bibliography{bibliography}

\end{document}